\documentclass[conference]{IEEEtran}
\IEEEoverridecommandlockouts
\usepackage{cite}
\usepackage{amsmath,amssymb,amsfonts}
\usepackage{algorithmic}
\usepackage{graphicx}
\usepackage{textcomp}
\usepackage{xcolor}
\usepackage{amsmath}
\usepackage{comment}
\usepackage{float}
\restylefloat{table}
\usepackage{threeparttable}
\usepackage{tabularx}
\usepackage{booktabs}

\usepackage{cite}

\def\BibTeX{{\rm B\kern-.05em{\sc i\kern-.025em b}\kern-.08em
    T\kern-.1667em\lower.7ex\hbox{E}\kern-.125emX}}

\begin{document}

\title{Fairness and Privacy in Federated Learning and Their Implications in Healthcare}

\author{\IEEEauthorblockN{Navya Annapareddy$^{1,*}$, Jade Preston$^{1,*}$, Judy Fox}
\IEEEauthorblockA{{$^{1}$School of Data Science}
University of Virginia, Charlottesville, VA \\
$^{*}$Authors contributed equally to this research\\
e-mail: \{na3au, yry2jy, ckw9mp\}@virginia.edu
}
}

\maketitle

\begin{abstract}
Currently, many contexts exist where distributed learning is difficult or otherwise constrained by security and communication limitations. One common domain where this is a consideration is in Healthcare where data is often governed by data-use-ordinances like HIPAA. On the other hand, larger sample sizes and shared data models are necessary to allow models to better generalize on account of the potential for more variability and balancing underrepresented classes. 

Federated learning is a type of distributed learning model that allows data to be trained in a decentralized manner. This, in turn, addresses data security, privacy, and vulnerability considerations as data itself is not shared across a given learning network nodes. Three main challenges to federated learning include node data is not independent and identically distributed (iid), clients requiring high levels of communication overhead between peers, and there is the heterogeneity of different clients within a network with respect to dataset bias and size. As the field has grown, the notion of fairness in federated learning has also been introduced through novel implementations. Fairness approaches differ from the standard form of federated learning and also have distinct challenges and considerations for the healthcare domain. This paper endeavors to outline the typical lifecycle of fair federated learning in research as well as provide an updated taxonomy to account for the current state of fairness in implementations. Lastly, this paper provides added insight into the implications and challenges of implementing and supporting fairness in federated learning in the healthcare domain.

\end{abstract}

\begin{IEEEkeywords}
Federated Learning, Healthcare, Medical Imaging, Fairness
\end{IEEEkeywords}

\section{Introduction}
Privacy and security have been a growing concern with Machine Learning systems. Having huge amounts of data in the same place is a security risk. Therefore, many domains such as healthcare regulate the centralization of data. Addressing this concern of centrally located data and privacy, McMahan et al. developed a concept called federated learning (FL). FL responds to the security issue of centralized information by having datasets at separate locations. It is a system where decentralized clients load and train data on their own devices and communicate their weights and predictions to a global server. Therefore, FL has been used to respond to security issues arising from clinical datasets. Though FL is prime for the incorporation of many devices, it has four challenges listed by McMahan et al. To start, the data in a FL framework is not independent and identically distributed because each device has different data types and represents a subset of the entire population. Additionally, some clients have more data than others resulting in an imbalance of information. Thirdly, there is an abundance of device or node connectivity - at times client participation is in the millions. Lastly, because of the large number of clients, communication between devices is difficult and the “cost” to the framework increases. \cite{McMahan}  \cite{Huang}\cite{lee2020federated}

\subsection{Fairness in Federated Learning}
Unfairness in FL can come from the way the clients are chosen, the methodology used to optimize the model, the way the clients are rewarded for contributions and the means by which the contributions to the model are defined. Typically, when choosing clients, speed and model accuracy are considered which can cause unfairness because some clients are always chosen, some are seldom chosen, and some are never chosen. Implementing fair FL techniques to individual clients can create an unfair federated model because of the lack of homogeneity of the clients and differences of data distribution for protected classes.\cite{Shi}\cite{Quy}

To combat unfairness within their FL frameworks many studies incorporate fairness tactics into their datasets or methodology; these tactics are driven by underlying ideas of how unfairness is brought into the framework. Some authors strive to ensure that each group within their system achieves the same or close to the same level or performance accuracy. Others seek to achieve fairer models through ensuring every group receives the same rewards and incentives no matter their contribution to the model. Continuing with this idea, some authors work to balance return of rewards for all groups. Rather than on the back-end of the study with rewards, others believe ensuring fairness begins with client submission to the study and make an  effort to motivate a diverse participation rate. Some authors believe that fairness is defined by the implementations within the algorithm; they work to modify weights or assumptions in the model setup to ensure every group receives appropriate metrics. Other researchers singularly strive to mitigate the loss within protected groups.\cite{Shi}

\subsection{Fair Federated Learning Assumptions}
Shi et al. outline nine different assumptions to fair FL: client data has privacy requirements, clients and servers are “self-serving”, logical, and honest, invitations to join FL are always accepted, clients always have prior information, and their data never changes, clients reserve all their resources for FL, and there is only one FL server operating. \cite{Shi}

\subsection{Federated Learning in Healthcare}
The healthcare domain is marked by special considerations for security and privacy that machine learning in healthcare (MLHC), including FL, must address. Specifically, most institutions subscribe to by ordinances that ensure patient privacy and confidentiality, commonly through compliance fixtures like the Healthcare Insurance Portability and Accountability Act (HIPAA). Even excluding such ordinances, it is generally preferable in healthcare contexts to isolate patient data from external access to preserve confidentiality and limit liability from unwanted access outside of clinical providers within a specific institution. \cite{ahmad2020fairness} This creates difficulty in conducting multi organizational or longitudinal studies as clinical information is siloed. This type of expansion is necessary as clinical studies often focus on a limited set of inclusion criteria leading to smaller sample sizes and a higher chance of carrying institutional bias if the data scope is limited.
Thus, FL for healthcare systems is proposed as a method to enable institutional collaboration and enable secure sharing of model parameters without compromising patient confidentiality. Notably, the resulting shared model weights are free of personally identifiable information (PII) as well as patient demographics. 
Implementations of FL in healthcare typically address the imbalanced nature of clinical patient samples as well as bias within different groups. Notably Quy et al. found significant demographic bias when performing a broad benchmark on clinical data ranging from signal data to medical imaging results. \cite{Quy}
Considerations for the use and fairness of FL in healthcare also must consider the secure environments within data exchanges must often occur. These can range from trusted execution environments (TEE) to secure encryption and handshake protocols that must be integrated into FL mechanisms for use in healthcare contexts. \cite{JAVAID202258}

\subsection{Gaps in Research}
There are four significant gaps within the research. Categorization of existing implementation are metric and resource allocation based. Fairness surveys focus on tabular data and ultimately lack considerations for image and high dimensional datasets, including ones commonly or exclusively found in healthcare such as signal and medical imaging datasets. Additionally, there has been no comprehensive lifecyle view to-date with an emphasis on fair FL. This makes it difficult to identify model process considerations or opportunities for security and privacy preservation across client nodes and servers. Lastly, there is no existing surveys with a specific focus on the relationship between fair FL implementations and its usage in healthcare context. 

Therefore the contributions of this paper will include: 1) an updated survey of current federated learning literature and implementations which incorporate fairness, 2) an updated taxonomy of fairness approaches resulting from the before mentioned survey, 3) a discrete mapping of a generalized FL model lifecyle process, and 4) provided insights on the healthcare implications drawn from the research completed in this paper.

\section{Related Work}
\subsection{Fairness Mechanisms in Federated Learning}
Recognizing the potential for improvements to fairness in ML using FL, many authors developed studies to address the concern. Shi et al developed a detailed classification for model fairness approaches. The outlined high level categories are as follows: client section, model optimization, contribution evaluation and incentive mechanism. \cite{Shi} These approaches as well as a few others will be expounded within this paper.

Quy et al. identified several fairness approaches that were not included in the taxonomy approach listed by Shi et al. such as  multi-discrimination, temporal fairness, and distributional fairness. In their paper ‘A survey on datasets for fairness-aware machine learning’ Quy et al. reviewed domain considerations of fairness. Specifically, they examine bias in FL datasets in the financial, health, and social domains. In each domain, the authors explored challenges to fairness such as data cardinality, imbalance, and the validity of assumptions of independence. Quy et al. concluded that: fairness-centric research defines fairness differently with every study/datasets, careful evaluation of protected classes needs to be conducted, and newer datasets need to be implemented to remain relevant. \cite{Quy}

We will later discuss how the following papers should be classified in terms of fairness approaches. Huang et al. authored a paper dealing specifically with FL and an incorporation of  fairness, ‘Fairness and Accuracy in Federated Learning’- introducing FedFa. FedFa is a FL optimization algorithm incorporating a double momentum gradient methodology and fairness weighting scheme. \cite{Huang} In another paper 'Ditto: Fair and Robust Federated Learning Through Personalization', Li et al. asserts that robustness and fairness are competing concepts in many FL frameworks. They assert that recent papers attempt to ensure fairness while also remaining robust, but don't realize that fairness and robustness are in fact competing concepts. By addressing this idea the authors introduce a methodology with an objective that makes FL more personalized.\cite{Ditto}
Hu et al. propose a different methodology for FL that adheres to group fairness requirements. They do this by expanding on the concept of bounded group loss (BGL). BGL is a methodology in which the average loss across all data points for each group are held the same. Continuing with this, the authors introduce an optimization formulation guaranteeing adherence to implemented group fairness requirements.\cite{BoundedGroupLoss}

\begin{figure*}
\begin{center}
\label{fig:fig1}
  \includegraphics[width=0.95\textwidth]{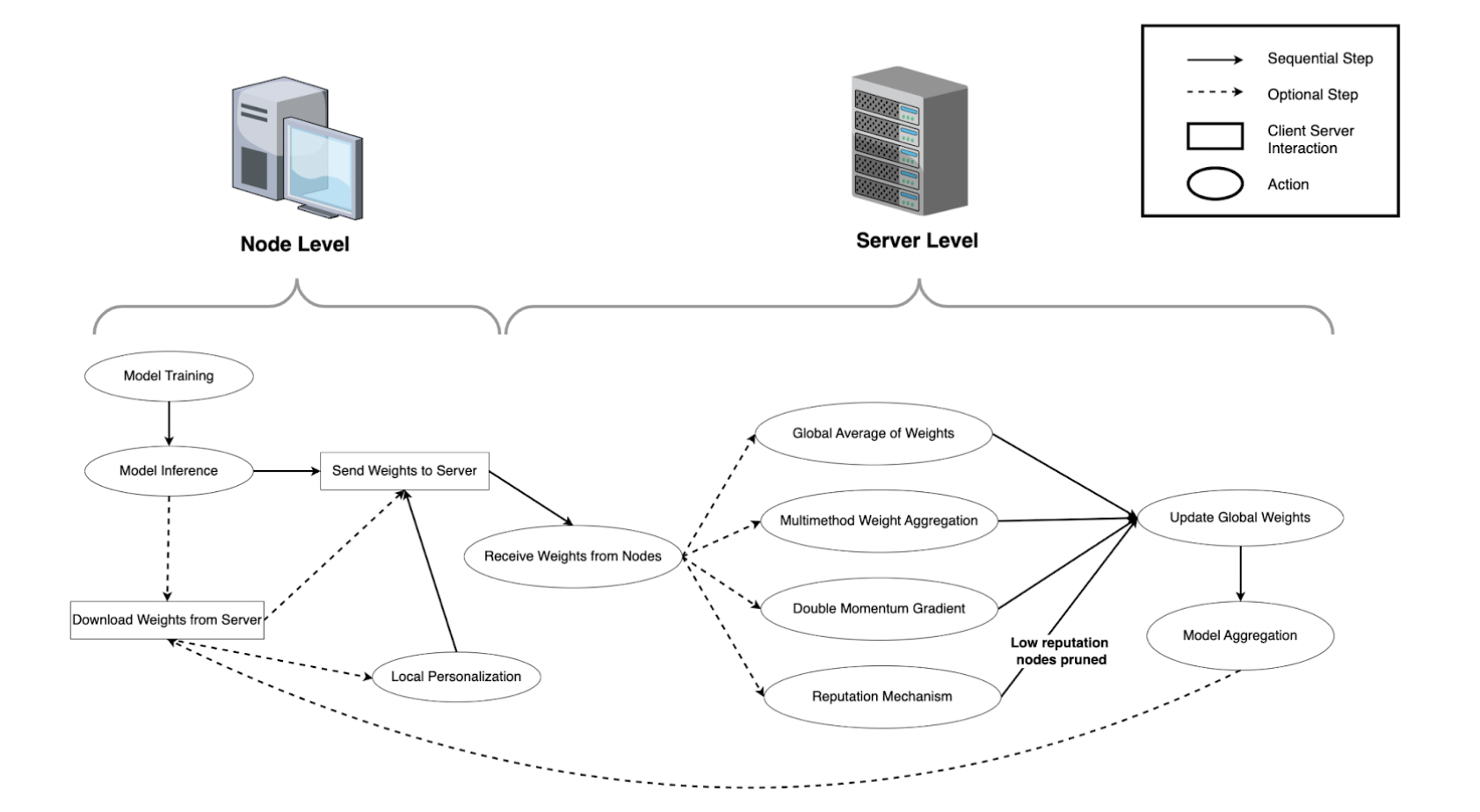}
  \caption{Discrete generalized federated learning lifecycle with client node level and server level processes and interactions. After model training and inference, key client server interactions include sending and downloading weights during aggregation steps.}
\end{center}
\end{figure*}

\subsection{Fairness in Context of Healthcare}
Fairness in healthcare is a broad concept that spans considerations for bias handling, representativeness, inclusion, generalizability, resource allocation, and disparity in results among the included groups. Specifically, given the history of algorithmic discrimination, MLHC must address both differential performance within and between patient groups as well as emergent fairness issues and appropriate responses.

Regulatory guidelines on defining and enforcing fairness within MLHC remain minimal. Notable guidelines include Standard Protocol Items: Recommendations for Interventional Trials-AI (SPIRIT-AI) and Consolidated Standards of Reporting Trials-AI (CONSORT-AI) which both mention encourage exploring bias as a consideration of model fairness but do not offer any format definition or response guidelines for doing so. \cite{Lium3164}

As engagement on fairness within MLHC remains preliminary, fairness within the context of FL in healthcare (FLHC) remains especially minimal.

\subsection{Security and Privacy in Healthcare}
The use of FL in healthcare has unique security and privacy considerations at each step of the model lifecycle. Trusted execution environments (TEEs) in healthcare are typically constrained in communication capabilities, especially between external organizations leading to the use of specialized servers and communication protocols to exchange model weights or enable client server interactions. \cite{JAVAID202258}

FLHC is also uniquely characterized by the participation of medical devices, wearable sensors, and Internet of Thing (IoT) enabled instruments. When learning infrastructure is split between client and server or multiple devices, the collaborative global model is vulnerable to attacks such as training-hijacking and targeted inversion attacks. \cite{sidechannel}

Differential privacy is highly relevant in healthcare contexts as it enables utilization of relevant features while minimizing available information on individuals in a group.

Finally, healthcare, as a domain, hosts a wide variety of datasets ranging from continuous signal inputs, spatio-temporal metadata, and emitting instruments with minimal encryption or privacy enabled features. The lack of standard usage for FLHC remains a challenge as privacy implementations remain context specific rather than generalizable.

\begin{figure*}
\begin{center}
\label{fig:fig2}
  \includegraphics[width=0.95\textwidth]{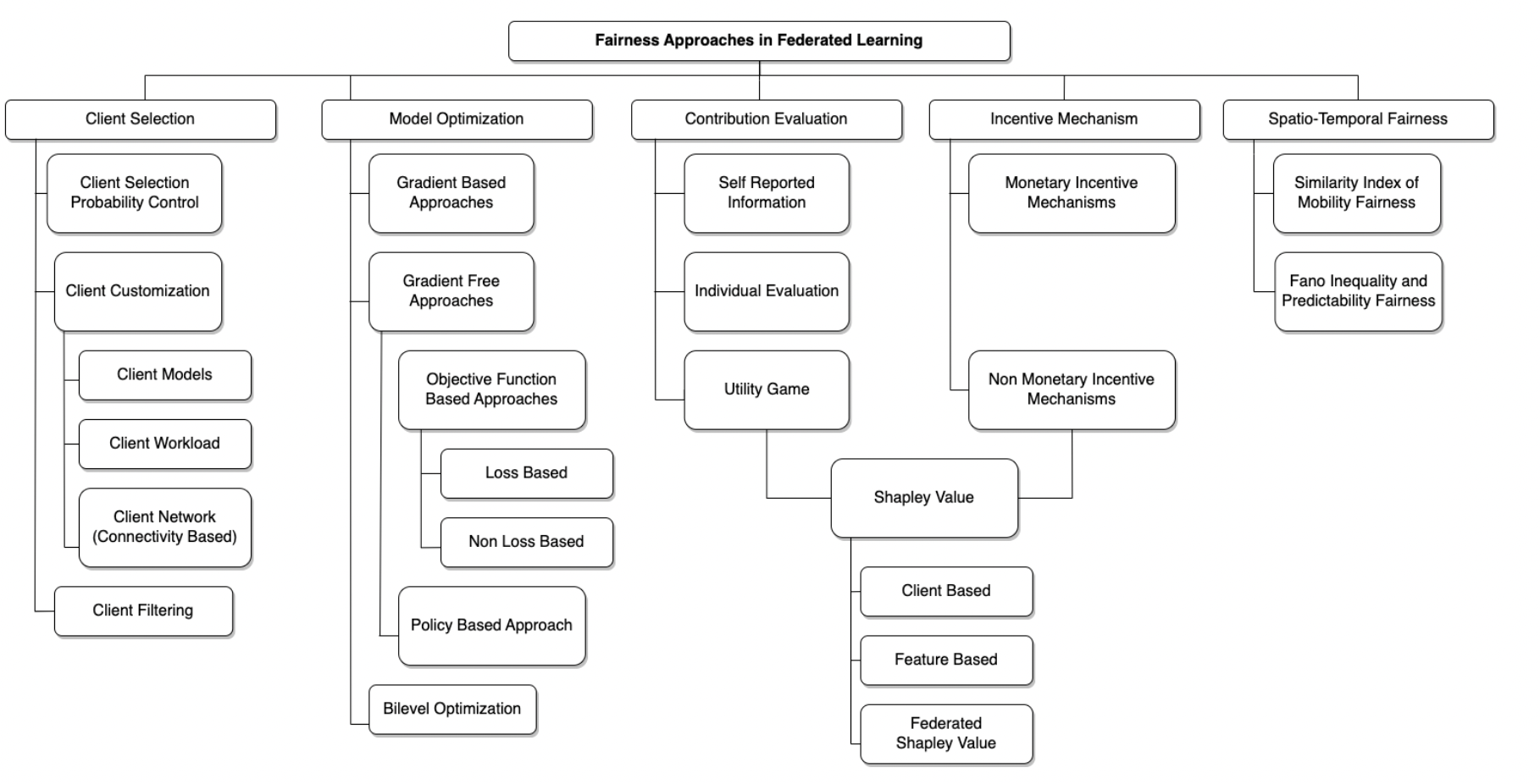}
  \caption{Generalized taxonomy of fairness approaches in federated learning (FAFL) detailing specific groups of high level classifications (client selection, model optimization, contribution evaluation, incentive mechanism, and spatio-temporal fairness) as well as each group's representative implementations.}
\end{center}
\end{figure*}

\section{Methodology}
To better understand the FL model process, a generalized view of the fairness FL lifecycle stages is depicted in Figure 1. The model representation consists of two stages of the process driven by the client nodes and global server respectively. Events known as client-server interaction (CSIs) and sequential events are present in each sub-process.

\subsection{Federated Learning Lifecycle}
\subsubsection{Client Level}
The lifecycle beginning at the node or client level begin very similarly across the studies we reviewed. Most federated frameworks have some form of model training at the node level. In ‘Fairness and Accuracy in Federated Learning’ by Huang et al., the authors implement a momentum gradient to train the model at the node level. They eventually incorporate a second momentum gradient a the server level resulting in faster convergence. \cite{Huang} 'Ditto: Fair and Robust Federated Learning Through Personalization' and 'Fair Federated Learning via Bounded Group Loss' are other examples of model training as part of the federated lifecycle. However, with these papers model training is conducted via minimizing an objective function. \cite{Ditto}\cite{BoundedGroupLoss} 

The next stage of the lifecycle at the node level usually involves the process of downloading weights. In 'A Reputation Mechanism Is All You Need: Collaborative Fairness and Adversarial Robustness in Federated Learning', the clients download the most recent global gradient and incorporate it to their local calculations. \cite{Reputation} Alternatively, with 'Ditto: Fair and Robust Federated Learning Through Personalization', a random sample of nodes are selected and are sent global weights from the server.\cite{Ditto} In addition to updating the weights,in 'Fair Federated Learning via Bounded Group Loss', the node receives a global fairness constraint from the server.\cite{BoundedGroupLoss}

The final overarching stage in the lifecycle is sending the updated weights to the server. No matter the methodology and in every study reviewed, the final process has weights sent from the local client level to the server. \cite{Ditto} \cite{BoundedGroupLoss} \cite{Reputation} An example of this stage can be observed in 'Federated Optimization in Heterogeneous Networks' by Li et al. In this paper, the authors modify a well known FL implementation FedAvg introduced in the original FL paper by McMahan et al. To accomplish this, Li et al. introduce FedProx and use an alternative local objective to minimize the impact of local updates and account for heterogeneous systems. FedProx assumes each device has their individual workload rather than the same. To account for the dissimilarity when weights are updated each round of FL, the authors incorporate variables into the algorithm. \cite{McMahan} \cite{FedProx} 

\subsubsection{Server Level}
FL schemes typically consist of transfer of local gradient and model updates from client nodes to a global server where aggregation steps and feedback loops between client and server can be enabled.

Model aggregation is a two-fold process with global aggregation of weights and/or gradients occurring before global models are ensembled and propagated back to clients. \cite{Quy}

Throughout this process, notions such as client contribution and participation are enforced. A given client's contribution can be minimized or maximized on the basis of a reputation objective function, as introduced by Xu et al. through their novel adversarial robustness metric \cite{Reputation}.

\subsection{Federated Learning Taxonomy}
We propose a comprehensive generalized FL taxonomy from relevant literature to expand current hierarchical understandings of the field to newer fairness mechanisms. In doing so, we will also expand the scope of previous reviews to text, image, and other high dimensional implementations commonly found in healthcare. We aim to differentiate the fairness approach and update the earlier taxonomy or pursue other organization methods based on developments in the field. In conjunction with a generalized FL lifecycle view, we review security and scaling considerations in the context of fairness, such as the security of metric aggregation, model sending, and scalability mechanisms such as peer selection and addition. The combined use of these discrete views help to expand past the assumption of single fairness mechanism systems and instead provide a basis to build a more complex, multi-fairness mechanisms for federated learning approaches. 

\section{Fairness Notions} 
\subsection{Performance parity fairness}
Shi et al. defines performance parity as the requirement of model accuracy equality for each group. This means that within the FL framework each group (protected or unprotected) will have a basic assured level of accuracy. \cite{Shi} A close comparison to this would be with the paper by Hu et al, 'Fair Federated Learning via Bounded Group Loss'. Their methodology aims to satisfy global model fairness requirements through saddle point optimization and Conditional Bounded Group Loss (CBGL). The idea of CBGL is specifically novel to this paper because it expounds on the concept BGL. With BGL, researchers ensure each group meets a minimum fairness baseline but does not require their fairness score to be equal. Hu et al. proposed BGL because 
 (by observing other similar techniques) they concluded that ensuring equal fairness scores causes a drop in prediction accuracy across all the groups. The authors state that CBGL is similar to the idea of equalized odds – true and false positive rate is made equal for each group. However, instead of requiring equality, CBGL sets an upper bound for each group. \cite{BoundedGroupLoss}

%

\subsection{Good Intent Fairness}
Good intent notions of fairness typically introduce concepts to prevent overfitting and minimize model bias, especially towards sensitive or protected data features. One interpretation of a model assuming good intent is one that minimizes losses for individual client nodes equally to avoid overfitting the model for the benefit of a subset of nodes, thereby minimizing bias to larger or more influential clients in the network.

One of the most significant implementations of good intent learning is the agnostic loss metric and corresponding agnostic federated learning (AFL) models.

Agnostic loss is a generalizable loss function able to optimize for a target distribution consisting of variable mixtures of clients, rather than a set distribution at runtime. \cite{afl} AFL models are more robust, while traditional FL schemes assume distribution parity between target and client distributions.  While this is true, AFL models have been found to be most effective for smaller FL networks and struggle to generalize accurately with large amounts of clients.

\subsection{Contribution Fairness}
Contribution fairness is a concept where the focus has shifted from ensuring a certain level of accuracy across groups to rewarding clients for their contribution to the model. \cite{Shi} Xu et. al focuses on two important concepts in 'A Reputation Mechanism Is All You Need: Collaborative Fairness and Adversarial Robustness in Federated Learning': collaborative fairness and adversarial robustness. Collaborative fairness is defined in the paper as the model’s ability to reward participants based on their individual and different contributions to the framework. Adversarial robustness is defined as the framework's ability to block “free riders” or malicious participants from reaping the benefits of the global model rewards. To implement collaborative fairness, Xu et al. reward participants based on the test accuracy of the client’s model. Adversarial robustness was addressed based on detection of model poisoning and the potential for free riders. \cite{Reputation}

\subsection{Regret Fairness}
In FL, regret fairness aims to minimize the difference between a node's expected input and the input it has actually received as well as the length at which the node has been in waiting for a given input. 
This notion relates to incentive schemes in that nodes follow an objective function that balances waiting and latency with incentive payout. \cite{Shi}

\subsection{Expectation Fairness}
Similar to regret fairness, the notion of expectation fairness aims to minimize inequity in client nodes as incentive payouts occur. Unlike regret fairness, expectation fairness assumes a gradual incentive payout such expectation values can be calculated. Since these payouts happen at different periods rather than all at once, it is useful for FL schemes where incentive budgets are not fixed or otherwise variable. \cite{expect}

\subsection{Spatio-temporal Fairness}
In ‘A survey on datasets for fairness-aware machine learning’, Quy et al. include the term temporal into their fairness notions. Specifically they discuss datasets with temporal and spacial information that should be removed to ensure the study is fair for all groups. Mashhadi et al. surveys the current approaches and challenges to fairness for special temporal frameworks in 'Fairness in Federated Learning for Spatial-Temporal Applications'. The authors assert that due to the rise of smart devices, research interest has progressed  in the study of trends in human mobility. Specifically, temporal and spatial fairness is a concern in response to the many problems with collecting large amounts of human mobility data in often unsecure data environments. There are privacy concerns to location information; this makes people less willing to share or upload their location. Second, the data repositories have missing or outdated information. Lastly, many of the datasets are discriminatory which would make any algorithm output biased toward certain groups. \cite{Quy}\cite{Spatial-Temporal}

\section{Fairness Approaches} 
Approaches to FAFL can also be classified discretely to standardize groups of implementations separate to fairness notions. This taxonomy depicted in Figure 2 outlines a hierarchy for the major schools of applications of fairness in federated learning as well as characteristic or relevant implementations.

\subsection{Client Selection}
In an attempt to incorporate fairness Shi et al. suggests there are  multiple approaches that fall into key categories. The first approach they classify is client selection. With client selection, nodes are allowed into the framework based on their potential. \cite{Shi} We assert that client selection is more accurately specified through additional specifications for client filtering and has strong ties to contribution evaluation through baseline constraints of accuracy or speed. With these contribution constraints, many clients are filtered out before the first round. Clients can also not only be rewarded but also removed every round of the study depending on their model performance. \cite{Reputation} 

\subsection{Incentive Mechanisms}
Typical incentive schemes in FL distribute rewards (monetary and otherwise) to clients in a standardized manner. This form does not indicate the vast differences client nodes may have in terms of local computational and physical resources (e.g. memory, bandwidth, battery power, and communication capacity and speed) \cite{incentive}

Because the standard form of an incentive scheme ignores that variation and inconsistency in local
updates of clients can affect the characteristics of the global model, it mimics the free-rider problem and prevents high quality contributors from being represented.

\subsection{Model Optimization}
Model optimization focuses on making changes to the algorithm to introduce fairness to the framework. \cite{Shi} Often the changes incorporated are the weights or some additional parameter. A great example of this concept is in 'Fairness and Accuracy in Federated Learning'. The authors Huang et al. modify the the FedAvg algorithm developed by McMahan et al. to generate FedFa.  In their methodology, Huang et al. update the algorithm from employing stochastic gradient decent. They incorporate momentum into the gradient at both the system node and server level to speed up convergence. The second modification to FedAvg was the weighting system. With FedFa, weights were updated and applied every round of the FL. To obtain the weights a random subset of the clients was selected; the weights were then calculated by a combination of the clients training accuracy and the number of times the client was selected. \cite{McMahan} \cite{Huang}

\subsection{Contribution Evaluation}
FL approaches that consider contribution evaluation typically consider fairness in the node selection, client filtering, and reward distribution stages. This approach is also key for differential privacy, as information that exposes individual level features can be removed for the sake of de-identification. \cite{contribution}

Contribution evaluation procedures are often based on features such as self reported sparsity, size, and weight improvements.

\section{Discussion: Implications In Healthcare} 
The use of fair FL in healthcare is governed by existing data usage and execution practices. While FL has the opportunity to increase in organizational collaboration by offering a secure channel for distributed weight sharing, it also has unique security challenges and overhead in communications. 

Client selection is commonly a function of nodes receiving weights for their potential additives to the model and different clients from different locations will thus provide data features characteristic to their area or group's demographics. Having a multitude of clients for various places will contribute to model accuracy for more groups within healthcare. Moreover, healthcare networks face frequent threats making the additional feedback loops and client server interactions (CSIs) necessary to enable aggregation methods like double momentum, local personalization, and reputation metrics less ideal than other approaches that minimize CSIs. On the other hand, these approaches incorporate essential elements of differential privacy by unburdening a given FL network from free rider or unhelpful clients and thus minimizing the use and propagation of unnecessary information.

Finally, an increasing amount of individual-level healthcare data is now sourced from wearable devices, sensors, or cloud enabled devices and smartphones that are responsible for tracking location and longitudinal health data. \cite{JAVAID202258} These edge devices are often less secured than traditional data collection devices and prone to vulnerabilities of users' local networks which is further compounded by lack of standardized use and security norms.

Overall, FLHC will be enabled by further consideration low resourced edge devices, minimization of communication overhead and organizational silos, and standardization of client and peer selection practices when considering FL in the context of existing biases and fairness implications in healthcare.

\section{Conclusion}
In this paper, we present a survey on the current state of fairness in federating learning as it pertains to healthcare. FAFL, especially in the healthcare domain, is not a zero-sum and as fairness approaches in federated learning evolve, approaches will converge to include more complex or multi-notion implementations. We expanded on existing taxonomy representations of FAFL by adding new fairness notions and considering implementations in the context of multiple categories and high level methodology groups. We also noted new FL paradigms such as bilevel optimization; this optimization is distinct from standard FL workflows in their continuous feedback loops between client and server. Finally, in this paper we also develop a discrete lifecycle representation of FL which can be paired with domain and execution environment knowledge to further model implementation and security considerations.

FLHC is a natural extension of the standard FL architecture and is aligned with the healthcare domain's focuses and ordinances pertaining to security, individual's privacy,  confidentiality, and group fairness. Applying FL to healthcare also brings many challenges to federated learning including high dimensionality, high communication overhead and complexity, and enforced security standards that may impair or otherwise complicate normal FL communication procedures. These challenges will continue to be relevant as the field of federated learning, fairness implementations, and their use in healthcare evolve.  

\section*{Acknowledgment}
The authors would like to thank the School of Data Science at the University of Virginia for their support during this research and providing access to publication catalogues. The authors also acknowledge the Research Computing at UVA for providing technical support that have contributed to the results reported within this publication.

\section*{Referenced code}
All the referenced and supporting code and survey iterations will be stored within the corresponding GitHub repository (https://github.com/UVA-MLSys/DS7406/tree/main/Projects/Team\%204).

\bibliographystyle{ieeetr}
\bibliography{references.bib}

\end{document}